\documentclass{elsart}
\usepackage{natbib}
\usepackage{epsfig}
\newcommand{\Msolar}{\mbox{\,$\rm M_{\odot}$}}        
\begin{document}
\runauthor{Cicero}
\begin{frontmatter}
\title{Supermassive black holes in radio-loud AGN}
\author[ross]{Ross J. McLure}
\address[ross]{Institute for Astronomy, University of Edinburgh}

\begin{abstract}
In this review the correlations between host galaxy properties and
black-hole mass determined from nearby quiescent galaxies are briefly
discussed, before proceeding to investigate their application to
active galactic nuclei (AGN). The recent advances in estimating the
black-hole masses of AGN are then reviewed, followed by an examination
of the connection between black-hole mass and radio luminosity.
\end{abstract}
\end{frontmatter}

\typeout{SET RUN AUTHOR to \@runauthor}

\section{Introduction}
Over the last decade the combination of HST and ground-based data has
led to the measurement of the black-hole masses of some 40 inactive
galaxies in the nearby Universe via gas and stellar
dynamics. These black-hole mass measurements have led to the discovery of 
correlations between black-hole mass and the properties of 
the surrounding host galaxy, namely bulge luminosity and 
stellar velocity dispersion. In parallel to the discovery of these
correlations in inactive galaxies, the last few years has also seen a new
industry emerge in estimating the black-hole masses of active
galaxies. In the first section of this review I will discuss the 
recent advances in our understanding of the connection 
between black-hole mass and
host-galaxy properties in both active and inactive galaxies at low 
redshift. In the second section I will proceed to discuss how 
black-hole mass estimation in active galaxies has been used to 
investigate the connection between black-hole mass and radio power.

\section{The black-hole masses of nearby inactive galaxies}
Within the context of the  study of AGN black-hole masses, the most 
important aspect of the recent increase in the numbers of
nearby quiescent galaxies with black-hole mass measurements has
been the discovery of two correlations between black-hole mass and
the properties of the surrounding host galaxy. Given the recent
demonstration that at low redshift the host galaxies of 
all powerful AGN are indistinguishable from nearby 
inactive ellipticals (Dunlop et al. 2003), irrespective of radio 
luminosity, these correlations can now be applied to estimate 
the central black-hole masses of active galaxies.

The first of these correlations is between black-hole mass and the
bulge luminosity of the host galaxy (Kormendy \& Richstone 1992,
Magorrian et al. 1998). This $M_{bh}-M_{bulge}$ correlation has the 
advantage of being observationally inexpensive, particularly 
with respect to radio galaxies, but is traditionally thought to 
suffer from a large associated scatter $\sim 0.5$ dex (although see below). 

The second, more recently discovered, correlation is between 
black-hole mass and the velocity dispersion of the host galaxy stellar
population (Gebhardt et al. 2000, Ferrarese \& Merritt 2000). This
correlation has been hailed as being the more ``fundamental'' due to
its very low associated scatter ($\sim 0.3$ dex). However,
the $M_{bh}-\sigma$ correlation, although extremely tight, has 
several disadvantages
with respect to studying the black-hole masses of active
galaxies. Firstly, even at low redshift it is not possible to 
obtain central stellar velocity dispersions for powerful quasars due 
to the presence of the unresolved nuclear point source. Furthermore, 
although it is possible to obtain
velocity dispersion measurements for radio galaxies at low redshift, 
this will not be practicable at $z>1$ even 
with 8m-class telescopes. Consequently, before proceeding to discuss a
more direct method of estimating black-hole mass in AGN, I will
digress slightly to discuss the possibility that the intrinsic scatter
in the $M_{bh}-M_{bulge}$ relation is significantly lower than has
previously been appreciated.

\subsection{The intrinsic scatter in the $M_{bh}-M_{bulge}$ relation}
\begin{figure}
\centerline{\epsfig{file=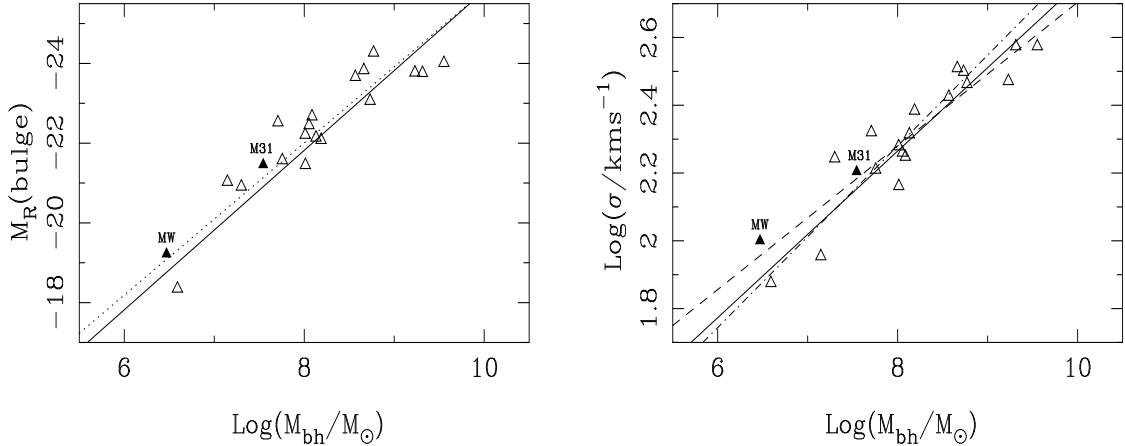,width=15cm,height=6cm,angle=0}}
\caption{A comparison of the $M_{bh}-M_{bulge}$ and $M_{bh}-\sigma$
relations for nearby inactive galaxies with pure E-type
morphologies. The scatter around the best-fitting relation in both
panels (solid lines) is 0.3 dex. The dashed and dot-dashed lines in
the right-hand panel are the $M_{bh}-\sigma$ fits of 
Ferrarese \& Merritt (2000) and Gebhardt et al. (2000) respectively. 
Figure taken from McLure \& Dunlop (2002).}
\label{fig1}
\end{figure}

It has occurred to a number of authors that the high level of scatter
associated with the $M_{bh}-M_{bulge}$ relation could well be an 
artifact of the difficulties in determining accurate bulge
luminosities for late-type galaxies. This issue is addressed in Fig
\ref{fig1} (taken from McLure \& Dunlop 2002) which shows a comparison of the
$M_{bh}-M_{bulge}$ and $M_{bh}-\sigma$ relations for a sample of 18
quiescent galaxies drawn from the list of objects with reliable
black-hole mass measurements published by Kormendy \& Gebhardt
(2001). These 18 objects comprise those objects from the Kormendy \&
Gebhardt compilation which have pure E-type morphology, and therefore
allow an investigation of the influence of incorrectly
determined bulge luminosity upon the scatter in the $M_{bh}-M_{bulge}$
relation. The best-fitting relations are shown as solid lines in 
both panels of Fig \ref{fig1} and display an identical level of 
associated scatter; $\sim 0.3$ dex. This comparison demonstrates that when
objects for which it is problematic to obtain accurate bulge
luminosities are excluded, the scatter in the $M_{bh}-M_{bulge}$ 
relation is comparable to that of the $M_{bh}-\sigma$ relation. 
This fact has obvious consequences for the study of black-hole mass in
radio-loud AGN, particularly with respect to radio galaxies for which it is
straightforward to obtain relatively accurate bulge luminosities
(eg. Bettoni et  al. 2003). Furthermore, Graham et al. (2001) and 
Erwin et al. (2002) have recently demonstrated that, at least 
for quiescent galaxies, if sufficient care is taken over the 
modelling of the galaxy surface brightness profiles, then the scatter
in both the $M_{bh}-M_{bulge}$ relation and a correlation between
black-hole mass and galaxy light concentration is only $\simeq 0.3$
dex, irrespective of morphological type.

\section{Estimating black-hole masses of low redshift AGN}

Although the black-hole mass - host galaxy correlations discussed
above are useful for estimating the black-hole masses of AGN, there is
a more direct method of
black-hole mass estimation which can be applied to broad-line AGN.

This method is the so-called virial black-hole mass estimator, and 
 utilizes the width of broad emission lines in quasar spectra 
to directly trace the gravitational potential of the 
central black-hole on the scale of light-days. The underlying
 assumption behind the method is that the motion of the
line-emitting material is virialized. Under this assumption the width
of the broad lines can be used to trace the Keplerian velocity
of the broad-line gas, and thereby allow an estimate of the
central black-hole mass via the formula :
 $M_{bh}=G^{-1}R_{BLR}V_{BLR}^{2}$, where $R_{BLR}$ is the broad-line
 region (BLR) radius 
and $V_{BLR}$ is the Keplerian velocity of the BLR gas. Recent
 evidence to support the
Keplerian interpretation of AGN broad-line widths has been
presented by Peterson \& Wandel (2000) and Onken \& Peterson (2002). These
authors demonstrate, at least for several well studied Seyfert
 galaxies, that the FWHM of various emission lines follow the
$V_{BLR}\propto R_{BLR}^{-0.5}$ relation expected for Keplerian
motion. It is also worth noting that Peterson (2002) has 
recently demonstrated that the virial black-hole masses of
reverberation mapped Seyfert galaxies lie on the same $M_{bh}-\sigma$
relation as nearby inactive galaxies.

\subsection{The $M_{bh}-M_{bulge}$ relation in low redshift AGN}
\begin{figure}
\centerline{\epsfig{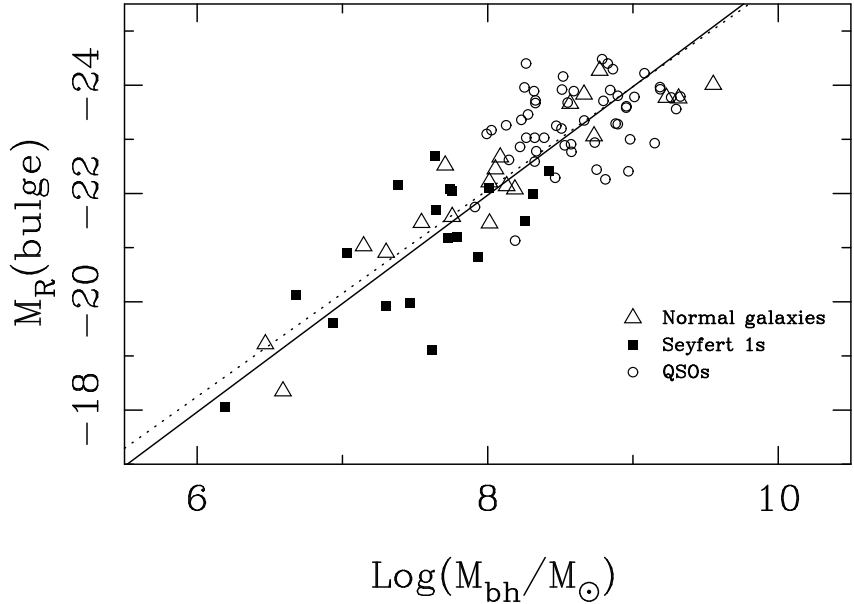}}
\caption{The $M_{bh}-M_{bulge}$ relation for a sample of 72
$z<0.5$ AGN and 18 nearby inactive galaxies with reliable
black-hole mass measurements. Quasars are shown as open circles,
Seyfert galaxies are shown as filled squares and inactive galaxies are
shown as open triangles. The black-hole masses of the AGN have been
estimated using the virial mass estimator assuming a disc-like BLR. 
The solid line is the best-fitting relation, while the dotted line is
the best-fitting relation with $M_{bh}$ and $M_{bulge}$ scaling
linearly. Figure taken from McLure \& Dunlop (2002)}
\label{fig2}
\end{figure}

By utilizing the virial black-hole mass estimator it is possible
to investigate whether the host galaxies of AGN do follow the same
$M_{bh}-M_{bulge}$ relation as inactive galaxies in the
nearby Universe (Laor 2001, McLure \& Dunlop 2002). In Fig \ref{fig2} 
the results of a study to investigate
the $M_{bh}-M_{bulge}$ relation of 72 $z<0.5$ AGN by McLure \& Dunlop
(2002) are shown. Also shown is the sample of nearby inactive
galaxies with reliable black-hole mass measurements previously
discussed in Fig \ref{fig1}. The AGN sample ($\sim 50$ \% radio-loud) 
consists of objects for which
the host-galaxy bulge luminosity was determined via modelling of
either HST or high-resolution ground-based imaging. It can clearly be
seen that the AGN host galaxies follow a tight (0.4 dex
scatter) correlation which is identical to that followed by nearby
quiescent galaxies. Indeed, the best-fitting linear relation (dotted
line) is equivalent to $M_{bh}=0.0012M_{bulge}$, identical to that
determined by Merritt \& Ferrarese (2001) for nearby quiescent galaxies.

\section{The radio power - black hole mass connection}
\begin{figure}
\centerline{\epsfig{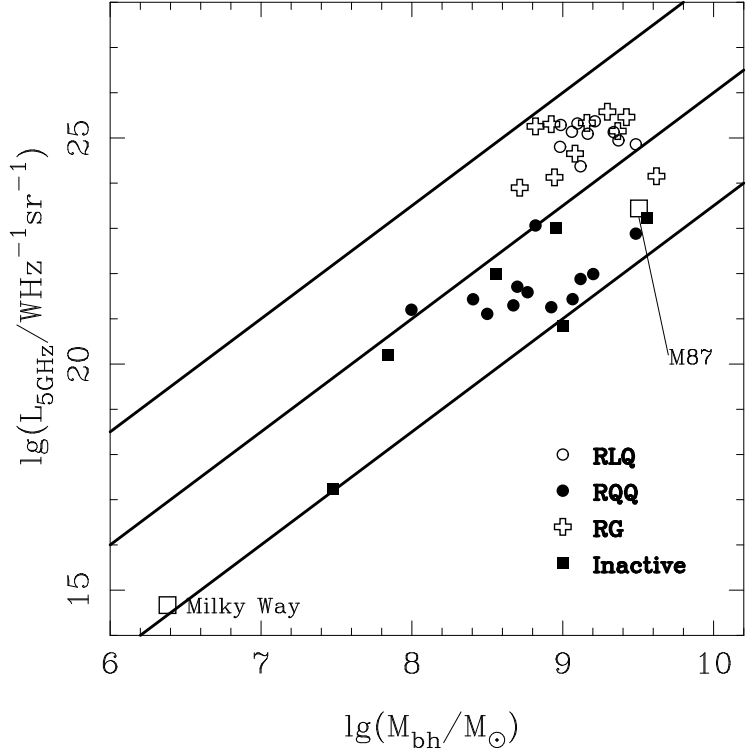}}
\caption{Total radio luminosity (5 GHz) versus black-hole mass for a sample
of 33 AGN from Dunlop et al. (2003) and nearby quiescent galaxies from
Franceschini et al. (1998). Figure taken from Dunlop et al. (2003).}
\label{fig3}
\end{figure}
In the second section of this review I will proceed to examine recent
evidence regarding the connection between black-hole mass
and radio power. Figure \ref{fig3} is taken from a recent paper by Dunlop
et al. (2003) and shows 5 GHz radio luminosity plotted against
black-hole mass for two samples. The first sample is comprised of 33
AGN, radio galaxies, radio-loud quasars and radio-quiet quasars, who's
bulge/black-hole masses have been determined from the analysis of HST
imaging. The second sample features the nearby quiescent galaxies 
investigated by Franceschini et al. (1998), who found a tight 
correlation between black-hole mass and 5 GHz radio power of the 
form: $P_{rad}\propto M_{bh}^{2.5}$. In Fig \ref{fig3} there are three
parallel relations plotted, each of the form  $P_{rad}\propto
M_{bh}^{2.5}$, separated from each other by 2.5 decades. The lowest of
these three relations appears to represent a lower limit to the 
radio output of a black-hole of a given mass.

There are two points concerning Fig \ref{fig3} which are worth 
highlighting. Firstly, it can be seen that many of the radio-quiet quasars 
from the Dunlop et al. sample lie up against the apparent radio power
lower limit. This is important because the 
radio-quiet and radio-loud quasars were selected by 
Dunlop et al. to have identical optical luminosities. 
Consequently, it can immediately be seen that the ``radio-quietness''
of these quasars is not due to their central engines being starved of
fuel, because they are still producing large optical luminosities. The 
second point refers to the hypothesized upper envelope plotted
in Fig \ref{fig3}. Clearly, if the Franceschini-type relation does 
represent a lower limit to the radio output of a given 
black-hole mass, then it is
interesting to inquire about the form of the corresponding upper
envelope. Although the form of the upper envelope
plotted in Fig \ref{fig3} is speculative, at least based on the 
data from Dunlop et al., further support is provided by the 
data from the study of Lacy et al. (2001).

\begin{figure}
\centerline{\epsfig{file=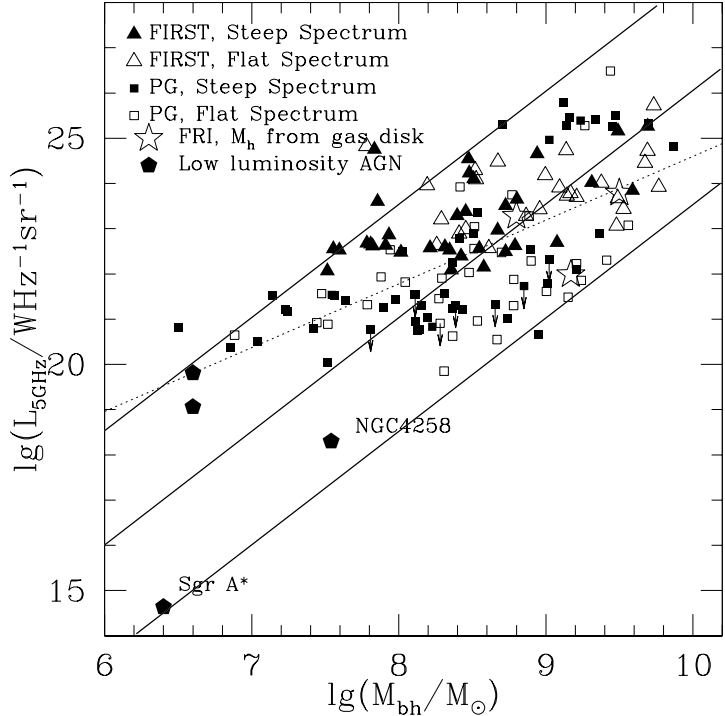,width=10cm,angle=0}}
\caption{Total radio luminosity (5 GHz) versus black-hole mass for the 
First Bright Quasar Survey (FBQS), the PG quasar survey and several 
notable low redshift objects. The black-hole 
masses of the quasars have been derived from the virial mass estimator. 
Figure adapted from Lacy et al. (2001)}
\label{fig4}
\end{figure}

Figure \ref{fig4} is an adapted version of Fig 2 from Lacy et al. (2001),
which again shows the $P_{rad}-M_{bh}$ plane, but this time populated by
objects from the First Bright Quasar Survey (FBQS), the PG quasar
survey and several notable low redshift objects. Figure \ref{fig4} 
clearly demonstrates that radio power and
black-hole mass are related, although a large dynamic
range is required in both parameters for the relation to become apparent. 

The dotted line in Fig \ref{fig4} is the best-fitting relation determined by 
Lacy et al. (not including accretion rate as an additional parameter) and has
the form $P_{rad}\propto M_{bh}^{1.4}$. However, in order to
offer an alternative interpretation of the Lacy et al. data I have 
annotated Fig \ref{fig4} with the addition of the upper
and lower limits on radio-power suggested by Dunlop et al. (2003). 
The lower radio power limit describes the Lacy et al. data extremely 
well, and again shows radio-quiet quasars lying tight up against the 
relation defined by quiescent galaxies, despite being well supplied 
with fuel. Furthermore, the Lacy et al. data also obey the upper limit
on radio-power over 4 orders of magnitude in black-hole
mass and 11 orders of magnitude in radio luminosity. Given that 
objects appear to exist between radio power boundaries
spanning 5 orders of magnitude, it is therefore understandable that 
studies which adopt the radio-loudness parameter $\mathcal{R}$ tend to find
little or no correlation between radio-loudness and black-hole mass
(eg. Woo \& Urry 2002).

One final interesting point is illustrated by Fig \ref{fig4}. Although 
there is clearly a large scatter associated with the
$P_{rad}-M_{bh}$ relation, it is still true that the most radio-loud
quasars are those with the most massive black-holes. For example, Fig
\ref{fig4} shows that quasars with $P_{rad}>10^{25}$WHzsr$^{-1}$ are 
virtually exclusively associated with black-holes of mass $>10^{9}
\Msolar$. One consequence of this
fact is that selecting samples based on extreme radio luminosity
offers a method for selecting objects with the largest black-holes,
and presumably bulge masses, at any given epoch. As a result, it
should therefore be possible to cleanly trace the evolution of the
$M_{bh}-M_{bulge}$ relation in a sub-set of the most massive galaxies
in the Universe.

\section{Conclusions}
The conclusions regarding the connection between black-hole mass and
radio power can be summarizes as follows:
\begin{itemize}
\item{There does not exist a threshold in black-hole mass above which
an AGN must be radio-loud. There is a large overlap in the black-hole
mass distributions of radio-loud and radio-quiet quasars, although at a
given optical luminosity radio-loud quasars do appear to be biased to
higher black-hole masses than their radio-quiet counterparts.}
\item{Black-hole mass and radio luminosity are connected, although an 
extremely large dynamic range in both parameters is required for this 
connection to become apparent.}
\item{The position of an object on the $P_{rad}-M_{bh}$ plane
appears to be a function of an additional physical parameter. Although
accretion rate may be playing a role, the lack of a tight correlation
between radio and optical luminosity suggests that some other
parameter, such as black-hole spin, may be involved.}
\item{Selecting the most radio-loud objects, as defined by their radio
luminosity alone, is an effective method of isolating the objects with
the most massive black-holes, and presumably host galaxies, at any
epoch.}
\item{Selecting the most radio-loud objects as a
function of redshift provides an ideal method for cleanly studying the
evolution of the $M_{bh}-M_{bulge}$ relation within the sub-set of the
most massive galaxies.}
\end{itemize}

\end{document}